\documentclass[9pt,twocolumn,twoside]{osajnl}

\journal{ol} 

\setboolean{shortarticle}{true}

\title{Platicon Stability in Hot Cavities}

\author[1,*]{Valery~E.~ Lobanov}
\author[2]{Nikita~M.~ Kondratiev}
\author[1]{Artem~E.~ Shitikov}
\author[3]{Olga~V.~ Borovkova}
\author[2]{Steevy J. Cordette}
\author[1,3]{Igor~A.~Bilenko}

\affil[1]{Russian Quantum Center, 143026, Skolkovo, Russia}
\affil[2]{Directed Energy Research Centre, Technology Innovation Institute, Abu Dhabi, United Arab Emirates}
\affil[3]{Faculty of Physics, Lomonosov Moscow State University, 119991, Moscow, Russia}

\affil[*]{Corresponding author: v.lobanov@rqc.ru}

\begin{abstract}
The stability of platicons in hot cavities with normal group velocity at the interplay of Kerr and thermal nonlinearities was addressed numerically.
The stability analysis was performed for different ranges of pump amplitude, thermal nonlinearity coefficient and thermal relaxation time. It was revealed that for the positive thermal effect, the high-energy wide platicons are stable, while the negative thermal coefficient provides the stability of narrow platicons. 

\end{abstract}

\setboolean{displaycopyright}{true}

\begin{document}

\maketitle
Over the past decades, the high-Q optical microresonators have proven to be an ideal platform for implementing, investigation, and application of various nonlinear optical phenomena \cite{Matsko2006, Strekalov_2016, Lin:17}. Numerous studies have shown that the dynamics of nonlinear processes in optical microresonators pumped by continuous wave lasers are greatly affected by the thermal effects, such as thermo-optic and thermal expansion effects \cite{Ilchenko1992ThermalNE, Fomin:05, Carmon:04}. The influence of these effects is almost inevitable in microresonator-based platforms due to the high field density inside the microresonator. It manifests itself as various thermal drifts, fluctuations, and instabilities \cite{Carmon:04, Diallo:15, PhysRevA.103.013512}. Their impact is especially strong and significant for such widely studied phenomena, as a generation of optical frequency combs and an excitation of the localized dissipative nonlinear states \cite{herr2014temporal, PASQUAZI20181}.

The thermal effects have been well-studied for the bright solitons that can be excited in the microresonators at anomalous group velocity dispersion (GVD) \cite{herr2014temporal, Bao:17, PhysRevLett.121.063902}. To decrease the impact of thermal effects many methods have been elaborated, e.g. various schemes of pump frequency sweep and pump power modulation \cite{Wildi:19, Li:17, Brasch357, Brasch:16}, active feedback circuits \cite{Yi:16} or even usage of cryogenic temperatures \cite{PhysRevApplied.12.034057} and additional laser \cite{Zhang:19, Zhou2019}. Another possible way to suppress the thermal effects during the frequency comb generation is a self-injection locking when pump frequency is locked to the microresonator eigenfrequency \cite{KondratievNum:20}.

In contrast to bright solitonic states, for dark solitons or platicons existing at normal GVD \cite{Lobanov2015, PhysRevA.89.063814}, the thermal effects and their impact have been studied in less detail. However, it is of great importance for many practical applications to understand complex nonlinear dynamics in microresonator-based photonic platforms at normal GVD at the interplay of Kerr and thermal nonlinearities. For example, it opens up the opportunities to estimate the operation area of the thermally induced generation of dark solitons that has been revealed recently \cite{Lobanov:21}. The proposed method of platicon generation doesn't require a complicated setup, such as a multi-resonator structure \cite{Xue2015, Kim:19} or photonic-crystal resonator \cite{yu2022continuum} for dispersion control  
or pump modulation systems \cite{Lobanov2019, Liu:22}. Moreover, the possibility of the turn-key regime of platicon generation enabled by the negative thermal effects was demonstrated. However, the applicability range of this approach is limited by the stability domain of platicons defined by the interplay of Kerr and thermal nonlinearities. 

In this Letter, we perform a numerical analysis of platicon stability in the presence of both Kerr and thermal nonlinearities for a wide range of parameters. It is fascinating for many practical applications since the thermal effects are unavoidable at the pump powers required for Kerr combs excitation. We found that high-energy wide platicons (or narrow dark solitons) are stable for the positive thermal effect, while the negative effect  stabilizes the narrow platicons.

We analyzed numerically the system of two equations: the Lugiato-Lefever equation \cite{PhysRevA.89.063814} for the slowly varying envelope of the intracavity field $\Psi$ and the rate equation for the normalized thermally induced detuning $\Theta$ \cite{Carmon:04,PhysRevA.103.013512,herr2014temporal,Diallo:15}:
\begin{gather}
    \label{ThermalLLE}
    \begin{cases}
        \frac{\partial\Psi}{\partial\tau}=\frac{d_2}{2}\frac{\partial^2\Psi}{\partial\varphi^2}-[1+i(\alpha-\Theta)]\Psi+i|\Psi|^2\Psi+f,\\
        \frac{\partial\Theta}{\partial\tau}=\frac{2}{\kappa t_T}\left(\frac{n_{2T}}{n_2}\frac{U}{2\pi}-\Theta\right). \\ 
    \end{cases}
\end{gather}
Here $\tau=\kappa t/2$ denotes the normalized time, 
$\kappa=\omega_0/Q$ is the cavity total decay rate ($Q$ is the loaded quality factor), 
$\omega_0$ is the pumped mode resonant frequency, 
$\varphi\in[-\pi;\pi]$ is an azimuthal angle in the coordinate frame rotating with the rate equal to the microresonator free spectral range (FSR) $D_1$, 
$d_2=2D_2/\kappa$ is the normalized GVD coefficient, positive for the anomalous GVD and negative for the normal GVD [the microresonator eigenfrequencies are assumed to be $\omega_\mu = \omega_0+D_1\mu+\frac{D_2}{2}\mu^2$ 
, where $\mu$ is the mode number, calculated from the pumped mode], 
$\alpha=2(\omega_0-\omega_p)/\kappa$ is the normalized detuning from the pump frequency $\omega_p$ from the pumped resonance.
The normalized pump amplitude for matched beam area and coupler refraction is $f=\sqrt{\frac{8\omega_{\rm p} c n_2\eta P_{\rm in}}{\kappa^2n^2 V_{\rm eff}}}$ \cite{herr2014temporal}, where $c$ is the speed of light, 
$n_2$ is the microresonator nonlinear index, 
$P_{\rm in}$ is the input pump power, 
$n$ is the refractive index of the microresonator mode, 
$V_{\rm eff}$ is the effective mode volume, 
$\eta$ is the coupling efficiency [$\eta=1/2$ for critical coupling, $\eta\rightarrow1$ for overloaded], $U=\int|\Psi|^2d\varphi$. 

The thermal relaxation time is not as important as its relation to the photon lifetime  $t_{ph}$  since $2/\kappa t_T = 2t_{ph}/t_T$. Parameters $t_T$  and $n_{2T}$ both depend on material properties and the geometry of the resonator, as a result their values can be significantly different for thermal refraction and thermal expansion \cite{Ilchenko1992ThermalNE, Diallo:15}. However, sometimes it is possible to neglect one of the effects, for example, when special composite structures are used for the compensation of thermal expansion \cite{Savchenkov_2018,Lim2017}. Thus, as a first step we consider a single composite effect.

First, we searched for stationary solutions of \eqref{ThermalLLE} (when $\frac{\partial\Psi}{\partial\tau} = 0$, $\frac{\partial\Theta}{\partial\tau} = 0$ ) in the form of dark solitons or platicons by means of the relaxation method using the following equation:
\begin{equation}
    \label{Stationary}
    \frac{d_2}{2}\frac{\partial^2\Psi}{\partial^2\varphi} - \left[1+i\left(\alpha-\frac{n_{2T}}{n_2}\frac{U}{2\pi}\right)\right]\Psi +i|\Psi|^2\Psi + f = 0. 
\end{equation}
Then the stability of obtained dark soliton solutions was tested by imposing random perturbations (additive and/or multiplicative) in the solutions and simulating their subsequent evolution up to $\tau \approx 10^4$  using \eqref{ThermalLLE}. Solitons that kept their shape during propagation were considered as the stable ones. 
Such analysis was performed for wide ranges of pump amplitude ($f = 1.5 ... 5$ ), normalized thermal relaxation time ($2/\kappa t_T = 0.001...10$), and for wide range of thermal nonlinearity coefficient both positive and negative ones ($-20 \le \frac{n_{2T}}{n_2} \le 20$). We set $d_2 = -0.02$ and checked that results were qualitatively the same for other values of the GVD coefficient.

\begin{figure}[t!]
\centering\includegraphics[width=1\linewidth]{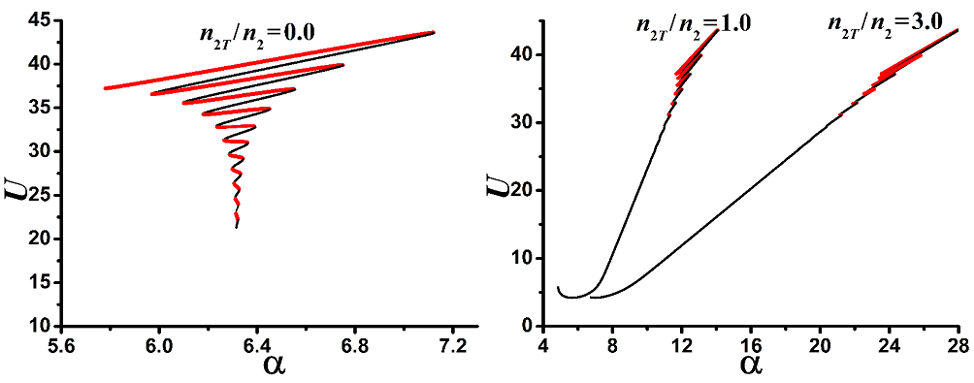}
\caption{\textbf{Left:} The platicon energy levels as the dependence of platicon energy $U$ on the pump frequency detuning $\alpha$ in the absence of thermal effects ($\frac{n_{2T}}{n_2} = 0$). Red lines correspond to stable platicons. \textbf{Right:} The platicon energy levels in presence of the positive thermal effects. Red lines correspond to stable platicons at $2/\kappa t_T = 0.1$. In all cases $f = 3$, $d_2 = -0.02$.
\label{fig1}}
\end{figure}

In thermally-compensated microresonators several discrete energy levels corresponding to stable dark solitons or platicons with different widths may exist in the same spectral range organized in a bifurcation structure known as collapsed snaking \cite{Lobanov2015,Parra-Rivas2016}. The dependence of platicon energy $U$ on the pump frequency detuning $\alpha$ is shown in the left panel of Fig. \ref{fig1} for $f = 3$, $\frac{n_{2T}}{n_2} = 0$. Stable solutions are indicated by red lines. Upper energy levels correspond to wider platicons. The stability domains become narrower with the decrease of the platicon width. 

Let's start with positive values of the thermal nonlinearity coefficient. In this case the thermal resonance shift had the same direction as nonlinear shift, we observed the shift, stretching and transformation of energy levels of platicons [compare left and right panels of Fig. \ref{fig1}]. Thermal nonlinearity partially lifts the degeneracy of platicon energy levels. However, it should be noted that at rather large values of thermal nonlinearity coefficient, the transition between different platicon levels by pump frequency scanning  is impossible: at forward scan with linear-in-time increase of detuning, the platicon transforms into low-intensity homogeneous state, while at backward scan the platicon comes into high-intensity homogeneous state.

Platicon stability was examined for different combinations of the pump amplitude, thermal nonlinearity coefficient and thermal relaxation time. It was revealed that for positive thermal nonlinearity only high-energy wide platicons from several upper energy levels can remain stable. Stability range becomes wider and shifts to the larger detuning values with the growth of the nonlinearity coefficient [see right panel in Fig. \ref{fig1}].

\begin{figure}[ht]
\centering\includegraphics[width=1\linewidth]{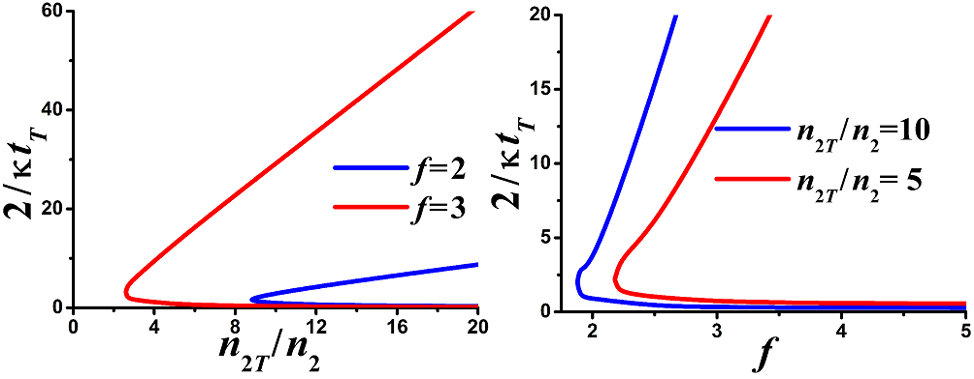}
\caption{The platicon stability domains on the thermal nonlinearity coefficient (left panel) and on the pump amplitudes (right panel). For the parameters indicated in the panels the platicons are unstable between solid lines of the corresponding color. In all cases $d_2 = -0.02$.
\label{fig2}}
\end{figure}

At the small values of the thermal nonlinearity coefficient the stability of platicon does not depend on the thermal relaxation time value. However, if the normalized thermal nonlinearity coefficient exceeds some threshold value, there appears a range of thermal relaxation time values providing platicon instability even for upper platicon levels. The dependence of the instability region on the pump amplitude $f$ and thermal nonlinearity coefficient $\frac{n_{2T}}{n_2}$ is shown in Fig. \ref{fig2}. It should be noted that the form of the collapsed snaking diagram does not depend on the thermal relaxation time as the \eqref{Stationary} does not depend on it.

\begin{figure}[ht]
\centering\includegraphics[width=1\linewidth]{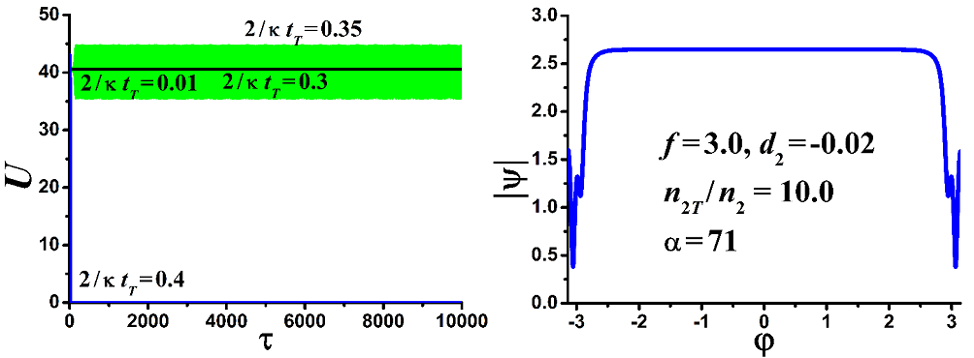}
\caption{\textbf{Left:} Scenarios of the intracavity power evolution for the platicon input and different values of thermal relaxation time at $\frac{n_{2T}}{n_2} = 10$, $f = 3$, $\alpha = 71$: stable platicon propagation at $2/\kappa t_T = 0.01$ and $2/\kappa t_T = 0.3$ –  solid black line, energy oscillations at $2/\kappa t_T = 0.35$ – green band, rapid platicon decay at $2/\kappa t_T = 0.4$ – blue line. \textbf{Right:} The input platicon amplitude profile.
\label{fig3}}
\end{figure}

The threshold value of $\frac{n_{2T}}{n_2}$, above which the unstable solutions appear, decreases with the growth of the pump amplitude [see left panel in the Fig. \ref{fig2}]. This instability range becomes wider with the growth of both thermal nonlinearity strength and the growth of the pump amplitude [see Fig. \ref{fig2}]. In the left panel of the Fig.~\ref{fig3} one can see three possible scenarios of the platicon dynamics. They were determined by propagation of the input amplitude profile like given in the right panel of the Fig.~\ref{fig3} perturbed by weak imposed noise. The stable solutions preserve their energy value as it is shown for two cases, $2/\kappa t_T = 0.01$ and $2/\kappa t_T = 0.3$. Close to the boundaries of the instability, region the platicons may experience energy oscillations (the green area), while far inside, as for $2/\kappa t_T = 0.4$, they may decay rapidly (the blue line in the plot that instantly falls down to zero).

The lower boundary of the stability range is more interesting since naturally thermal relaxation time is much larger than photon lifetime and $2/\kappa t_T \ll 1$. For a wide range of parameters, the platicons were found to be stable if $2/\kappa t_T \le 0.1$ -- that is true for the majority of the real-life microresonator platforms.

\begin{figure}[ht]
\centering\includegraphics[width=1\linewidth]{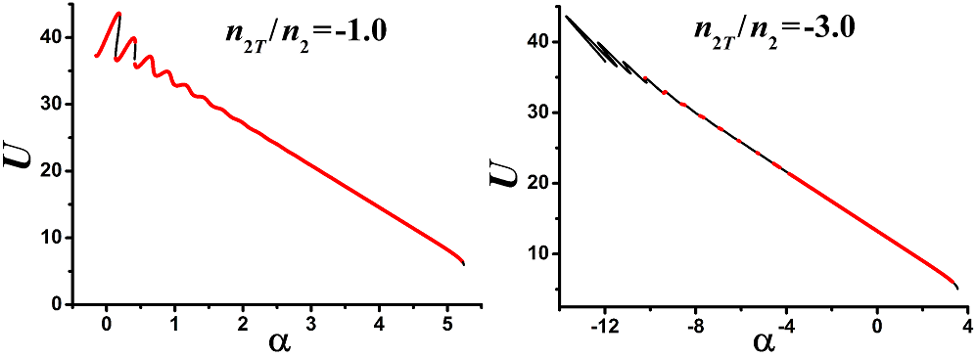}
\caption{The platicon energy levels in the presence of negative thermal effects. Red lines correspond to stable platicons at $2/\kappa t_T = 0.1$. In all cases $f = 3$, $d_2 = -0.02$.
\label{fig4}}
\end{figure}

\begin{figure}[ht]
\centering\includegraphics[width=1\linewidth]{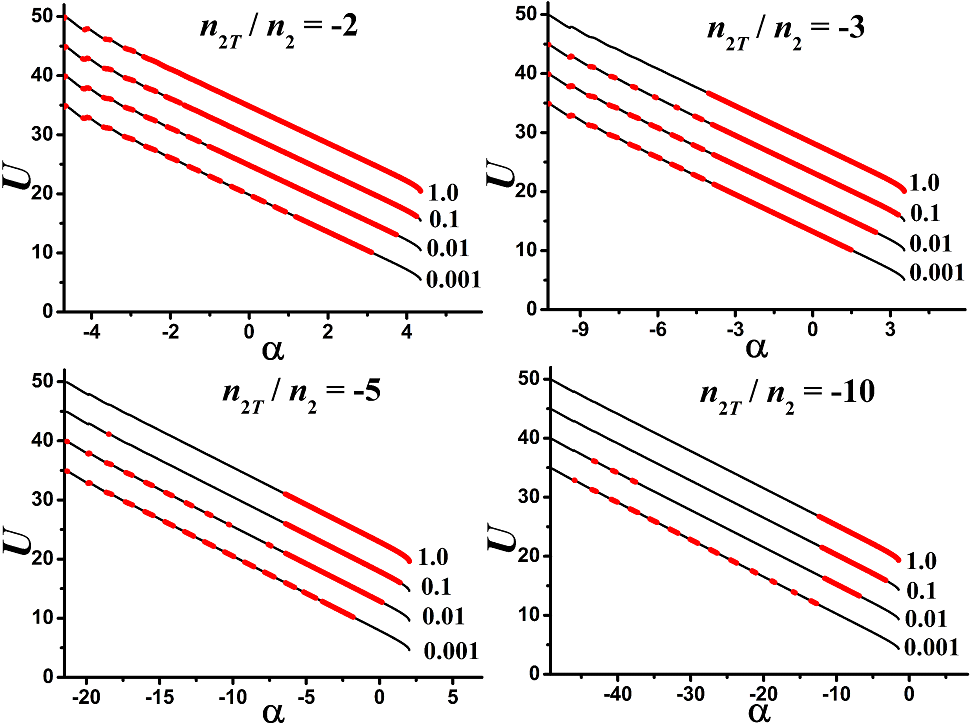}
\caption{Stability domains indicated by solid red lines for different values of $2/\kappa t_T$. For clarity, in each panel corresponding to the particular value of thermal nonlinearity coefficient several identical curves $U(\alpha)$ are shifted relative to each other along the vertical axis. The bottom curve for each panel is obtained by numerical solution of \eqref{Stationary}. The $2/\kappa t_T=$0.01, 0.1 and 1 curves are copies of $2/\kappa t_T=0.001$ one virtually shifted along vertical axes by 5, 10 and 15, correspondingly for clarity.
\label{fig5}}
\end{figure}

Now let's consider the second case of the negative thermal nonlinearity.
If $n_{2T} < 0$, then the stability conditions become significantly different. Instability arises if the thermal nonlinearity strength $|\frac{n_{2T}}{n_2}|$ exceeds the threshold value that decreases with the growth of the pump amplitude [see Figs.~\ref{fig4},~\ref{fig5}]. Below this threshold [at $|\frac{n_{2T}}{n_2} \le 1$ for $f = 3$], the high-energy platicons from the discrete levels and low-energy platicons from the continuous level are stable. 

With the growth of the thermal nonlinearity strength, high-energy wide platicons become unstable and most of low-power narrow platicons remain stable. The dependence of the platicon energy $U$ on pump frequency detuning $\alpha$ is shown in Fig. \ref{fig4} for $f = 3$ at $\frac{n_{2T}}{n_2} = -1.0$ (left panel) and at $\frac{n_{2T}}{n_2} = -3.0$ (right panel). Stable solutions are indicated by red lines. The structure of the spectral stability domains may be rather complicated and strongly depends on the normalized thermal relaxation time value or on the ratio of the thermal relaxation time and photon lifetime $2/\kappa t_T$. It is shown in the Fig. \ref{fig5} where each panel corresponds to the particular value of the thermal nonlinearity coefficient. For clarity several identical curves $U(\alpha)$ describing stability domains for different values of $2/\kappa t_T$ are shifted relative to each other along the vertical axis.

\begin{figure}[ht]
\centering\includegraphics[width=1\linewidth]{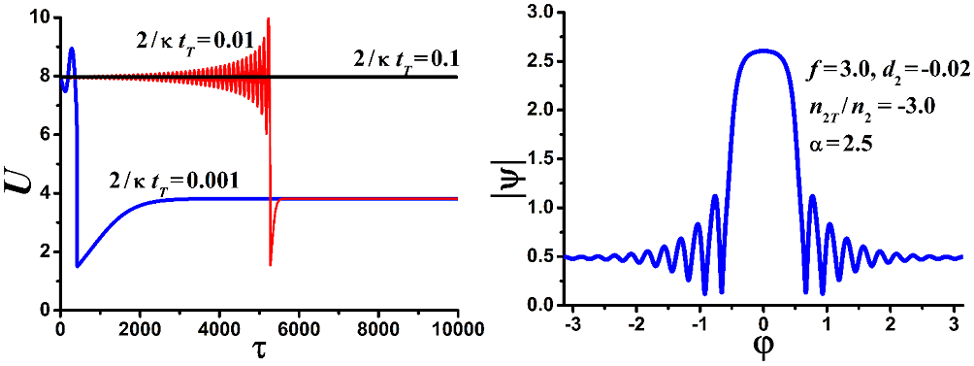}
\caption{\textbf{Left:} The scenarios of the intracavity power evolution for the platicon input and different values of the thermal relaxation time at $\frac{n_{2T}}{n_2} = -3.0$, $f = 3$, $\alpha = 2.5$: stable platicon propagation at $2/\kappa t_T = 0.1$  –  solid black line, unstable propagation and decay at $2/\kappa t_T = 0.01$ (red line) and $2/\kappa t_T = 0.001$ (blue line). \textbf{Right:} Input platicon amplitude profile.
\label{fig6}}
\end{figure}

\begin{figure}[ht]
\centering\includegraphics[width=1\linewidth]{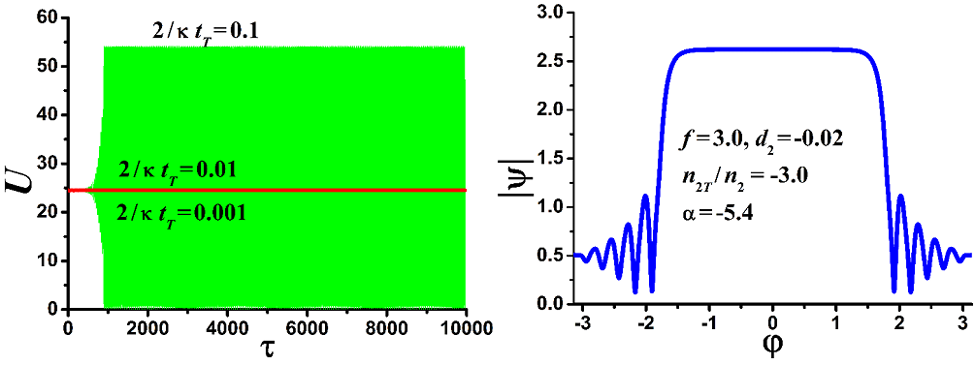}
\caption{\textbf{Left:} The scenarios of the intracavity power evolution for the platicon input and different values of the thermal relaxation time at $\frac{n_{2T}}{n_2} = -3.0$, $f = 3$, $\alpha = -5.4$: stable platicon propagation at $2/\kappa t_T = 0.01$ and $2/\kappa t_T = 0.001$ –  solid red line, unstable propagation and decay at $2/\kappa t_T = 0.1$ (red line) – green band. \textbf{Right:} Input platicon amplitude profile.
\label{fig7}}
\end{figure}

One may notice that the difference between structures of the stability domains shown by red lines for different values of the thermal relaxation time becomes more and more pronounced with the growth of the thermal nonlinearity strength [compare panels in Fig. \ref{fig5}]. If photon lifetime to thermal relaxation time ratio $2/\kappa t_T$ exceeds the critical value, then there is comparatively wide stability domain close to the right boundary of the existence domain. This domain becomes narrower with decrease of $2/\kappa t_T$. If $2/\kappa t_T$ is less than the critical value for this domain, then the platicon experiences a decay upon propagation. This is shown in Fig. \ref{fig6} for the parameters from top right panel in Fig. \ref{fig5}. Outside this widest stability domain there may be several narrow stability domains, but they can exist if $2/\kappa t_T$ is less than some critical value [see Figs. \ref{fig5} and \ref{fig7}]. Different scenarios of the evolution of the platicon from this domain are shown in Fig. \ref{fig7} for the parameters from top right panel in Fig. \ref{fig5}. Note, that with the further growth of the nonlinearity strength the upper threshold for $2/\kappa t_T$ also appears for this wide right stability domain. Upper threshold value decreases with the growth of $|\frac{n_{2T}}{n_2}|$ and increases with decrease of pump frequency detuning. For example, for $f=3$ and $\frac{n_{2T}}{n_2} = -3.0$ it varies from  $2/\kappa t_T=0.68$ at right boundary of the platicon existence domain ($\alpha=-8.29$) up to $2/\kappa t_T=2.19$ at the left boundary of the widest stability domain ($\alpha =-25$). 

\begin{figure}[ht]
\centering\includegraphics[width=1\linewidth]{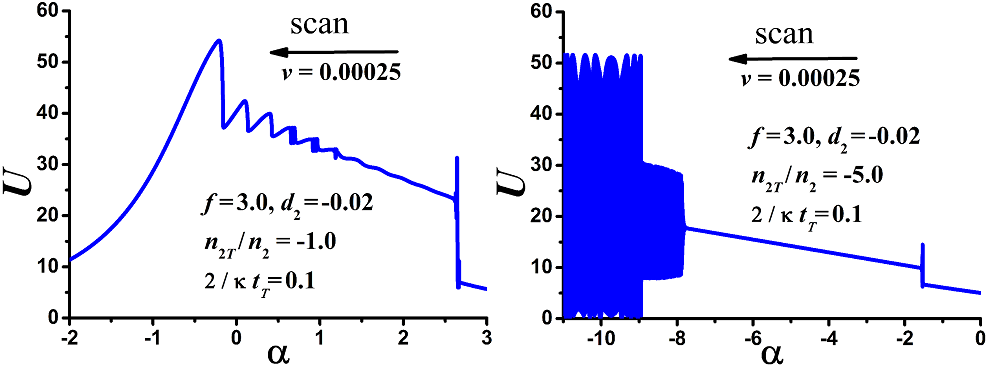}
\caption{The intracavity power evolution upon linear-in-time backward pump frequency scan $\alpha(\tau) = \alpha(0)-v\tau$ at $f=3$, $d_2=-0.02$, $2/\kappa t_T = 0.1$, $v=0.00025$  for $\frac{n_{2T}}{n_2} = -1.0$ (left panel) and $\frac{n_{2T}}{n_2} = -5.0$ (right panel). Platicon excitation occurs at first from the right power jump.
\label{fig8}}
\end{figure}

We have verified that the thermally induced generation of stable platicons \cite{Lobanov:21} upon forward scan (with linear-in-time increase of pump frequency detuning) occurs within the right stability domain. Moreover, it was revealed that the platicon generation is also possible upon backward scan ($\alpha(\tau) = \alpha(0)-v\tau$) that is shown in the Fig. \ref{fig8}. In that case the platicon tuning range is limited by the width of the right stability domain. One may see in the Fig. \ref{fig8} that at $f=3$, $d_2=-0.02$, $2/\kappa t_T = 0.1$, $v=0.00025$ generation of stable platicons takes place for $\frac{n_{2T}}{n_2} = -1.0$ (left panel, some power oscillations correspond to the transition between the platicon energy levels shown in the left panel of Fig. \ref{fig4}), while for $\frac{n_{2T}}{n_2} = -5.0$ platicons become unstable at some detuning value (right panel; corresponding stability domains are shown in the bottom left panel of Fig. \ref{fig5}). This also impose limitations on the maximum power of the stable platicon.

It should be noted that the spectral width of the stability range at the large negative thermal nonlinearity  coefficient is significantly smaller than for the same positive thermal effect value. For example, at $\frac{n_{2T}}{n_2} = -20.0$, $f = 3$  the normalized width of the stability domain for $2/\kappa t_T = 0.1$ is $\delta \alpha_{stab} \approx 10$, while for $\frac{n_{2T}}{n_2} = 20.0$ $\delta\alpha_{stab}  \approx 35$.

To sum up, we have analyzed the stability of platicons in hot cavities for different combinations of pump amplitude,
thermal nonlinearity coefficient and thermal relaxation time. It was revealed that if the thermal effects are positive, the high-energy wide platicons are stable; if thermal effects are negative, low-energy narrow platicons are mostly stable. The obtained results provide a deep insight in complex nonlinear dynamics in microresonator-based photonic platforms at normal GVD and can help to estimate the applicability range of the platicon generation method based on thermal effects.

\begin{backmatter}
\bmsection{Funding} 
RQC team was supported by the Russian Science Foundation (project 22-22-00872).
\bmsection{Acknowledgments} 
 V.E.L. and O.V.B. acknowledge the personal support from the Foundation for the Advancement of Theoretical Physics and Mathematics “BASIS”.
\bmsection{Disclosures} The authors declare no conflicts of interest.
\end{backmatter}

\bibliography{sample}

\begin{thebibliography}{10}
\newcommand{\enquote}[1]{``#1''}

\bibitem{Matsko2006}
V.~Ilchenko and A.~Matsko, {\protect\JournalTitle{IEEE Journal of Selected
  Topics in Quantum Electronics}} \textbf{12}, 15 (2006).

\bibitem{Strekalov_2016}
D.~V. Strekalov, C.~Marquardt, A.~B. Matsko, H.~G.~L. Schwefel, and G.~Leuchs,
  {\protect\JournalTitle{Journal of Optics}} \textbf{18}, 123002 (2016).

\bibitem{Lin:17}
G.~Lin, A.~Coillet, and Y.~K. Chembo, {\protect\JournalTitle{Adv. Opt.
  Photon.}} \textbf{9}, 828 (2017).

\bibitem{Ilchenko1992ThermalNE}
V.~Ilchenko and M.~L. Gorodetskii, {\protect\JournalTitle{Laser Physics}}
  \textbf{2}, 1004 (1992).

\bibitem{Fomin:05}
A.~E. Fomin, M.~L. Gorodetsky, I.~S. Grudinin, and V.~S. Ilchenko,
  {\protect\JournalTitle{J. Opt. Soc. Am. B}} \textbf{22}, 459 (2005).

\bibitem{Carmon:04}
T.~Carmon, L.~Yang, and K.~J. Vahala, {\protect\JournalTitle{Opt. Express}}
  \textbf{12}, 4742 (2004).

\bibitem{Diallo:15}
S.~Diallo, G.~Lin, and Y.~K. Chembo, {\protect\JournalTitle{Opt. Lett.}}
  \textbf{40}, 3834 (2015).

\bibitem{PhysRevA.103.013512}
A.~Leshem, Z.~Qi, T.~F. Carruthers, C.~R. Menyuk, and O.~Gat,
  {\protect\JournalTitle{Phys. Rev. A}} \textbf{103}, 013512 (2021).

\bibitem{herr2014temporal}
T.~Herr, V.~Brasch, J.~D. Jost, C.~Y. Wang, N.~M. Kondratiev, M.~L. Gorodetsky,
  and T.~J. Kippenberg, {\protect\JournalTitle{Nat. Photon.}} \textbf{8}, 145
  (2014).

\bibitem{PASQUAZI20181}
A.~Pasquazi, M.~Peccianti, L.~Razzari, D.~J. Moss, S.~Coen, M.~Erkintalo, Y.~K.
  Chembo, T.~Hansson, S.~Wabnitz, P.~Del’Haye, X.~Xue, A.~M. Weiner, and
  R.~Morandotti, {\protect\JournalTitle{Physics Reports}} \textbf{729}, 1
  (2018). Micro-combs: A novel generation of optical sources.

\bibitem{Bao:17}
C.~Bao, Y.~Xuan, J.~A. Jaramillo-Villegas, D.~E. Leaird, M.~Qi, and A.~M.
  Weiner, {\protect\JournalTitle{Opt. Lett.}} \textbf{42}, 2519 (2017).

\bibitem{PhysRevLett.121.063902}
J.~R. Stone, T.~C. Briles, T.~E. Drake, D.~T. Spencer, D.~R. Carlson, S.~A.
  Diddams, and S.~B. Papp, {\protect\JournalTitle{Phys. Rev. Lett.}}
  \textbf{121}, 063902 (2018).

\bibitem{Wildi:19}
T.~Wildi, V.~Brasch, J.~Liu, T.~J. Kippenberg, and T.~Herr,
  {\protect\JournalTitle{Opt. Lett.}} \textbf{44}, 4447 (2019).

\bibitem{Li:17}
Q.~Li, T.~C. Briles, D.~A. Westly, T.~E. Drake, J.~R. Stone, B.~R. Ilic, S.~A.
  Diddams, S.~B. Papp, and K.~Srinivasan, {\protect\JournalTitle{Optica}}
  \textbf{4}, 193 (2017).

\bibitem{Brasch357}
V.~Brasch, M.~Geiselmann, T.~Herr, G.~Lihachev, M.~H.~P. Pfeiffer, M.~L.
  Gorodetsky, and T.~J. Kippenberg, {\protect\JournalTitle{Science}}
  \textbf{351}, 357 (2016).

\bibitem{Brasch:16}
V.~Brasch, M.~Geiselmann, M.~H.~P. Pfeiffer, and T.~J. Kippenberg,
  {\protect\JournalTitle{Opt. Express}} \textbf{24}, 29312 (2016).

\bibitem{Yi:16}
X.~Yi, Q.-F. Yang, K.~Y. Yang, and K.~Vahala, {\protect\JournalTitle{Opt.
  Lett.}} \textbf{41}, 2037 (2016).

\bibitem{PhysRevApplied.12.034057}
G.~Moille, X.~Lu, A.~Rao, Q.~Li, D.~A. Westly, L.~Ranzani, S.~B. Papp,
  M.~Soltani, and K.~Srinivasan, {\protect\JournalTitle{Phys. Rev. Applied}}
  \textbf{12}, 034057 (2019).

\bibitem{Zhang:19}
S.~Zhang, J.~M. Silver, L.~D. Bino, F.~Copie, M.~T.~M. Woodley, G.~N. Ghalanos,
  A.~{\O}. Svela, N.~Moroney, and P.~Del'Haye, {\protect\JournalTitle{Optica}}
  \textbf{6}, 206 (2019).

\bibitem{Zhou2019}
H.~Zhou, Y.~Geng, W.~Cui, S.-W. Huang, Q.~Zhou, K.~Qiu, and C.~Wei~Wong,
  {\protect\JournalTitle{Light: Science {\&} Applications}} \textbf{8}, 50
  (2019).

\bibitem{KondratievNum:20}
N.~M. Kondratiev, V.~E. Lobanov, E.~A. Lonshakov, N.~Y. Dmitriev, A.~S.
  Voloshin, and I.~A. Bilenko, {\protect\JournalTitle{Opt. Express}}
  \textbf{28}, 38892 (2020).

\bibitem{Lobanov2015}
V.~E. Lobanov, G.~Lihachev, T.~J. Kippenberg, and M.~L. Gorodetsky,
  {\protect\JournalTitle{Opt. Express}} \textbf{23}, 7713 (2015).

\bibitem{PhysRevA.89.063814}
C.~Godey, I.~V. Balakireva, A.~Coillet, and Y.~K. Chembo,
  {\protect\JournalTitle{Phys. Rev. A}} \textbf{89}, 063814 (2014).

\bibitem{Lobanov:21}
V.~E. Lobanov, N.~M. Kondratiev, and I.~A. Bilenko, {\protect\JournalTitle{Opt.
  Lett.}} \textbf{46}, 2380 (2021).

\bibitem{Xue2015}
X.~Xue, Y.~Xuan, P.-H. Wang, Y.~Liu, D.~E. Leaird, M.~Qi, and A.~M. Weiner,
  {\protect\JournalTitle{Laser \& Photonics Reviews}} \textbf{9}, L23 (2015).

\bibitem{Kim:19}
B.~Y. Kim, Y.~Okawachi, J.~K. Jang, M.~Yu, X.~Ji, Y.~Zhao, C.~Joshi, M.~Lipson,
  and A.~L. Gaeta, {\protect\JournalTitle{Opt. Lett.}} \textbf{44}, 4475
  (2019).

\bibitem{yu2022continuum}
S.-P. Yu, E.~Lucas, J.~Zang, and S.~B. Papp, {\protect\JournalTitle{Nature
  Communications}} \textbf{13}, 1 (2022).

\bibitem{Lobanov2019}
V.~E. Lobanov, N.~M. Kondratiev, A.~E. Shitikov, R.~R. Galiev, and I.~A.
  Bilenko, {\protect\JournalTitle{Phys. Rev. A}} \textbf{100}, 013807 (2019).

\bibitem{Liu:22}
H.~Liu, S.-W. Huang, W.~Wang, J.~Yang, M.~Yu, D.-L. Kwong, P.~Colman, and C.~W.
  Wong, {\protect\JournalTitle{Photon. Res.}} \textbf{10}, 1877 (2022).

\bibitem{Savchenkov_2018}
A.~Savchenkov and A.~Matsko, {\protect\JournalTitle{Journal of Optics}}
  \textbf{20}, 035801 (2018).

\bibitem{Lim2017}
J.~Lim, A.~A. Savchenkov, E.~Dale, W.~Liang, D.~Eliyahu, V.~Ilchenko, A.~B.
  Matsko, L.~Maleki, and C.~W. Wong, {\protect\JournalTitle{Nature Commun.}}
  \textbf{8}, 8 (2017).

\bibitem{Parra-Rivas2016}
P.~Parra-Rivas, E.~Knobloch, D.~Gomila, and L.~Gelens,
  {\protect\JournalTitle{Phys. Rev. A}} \textbf{93}, 063839 (2016).

\end{thebibliography}

\bibliographyfullrefs{sample}

\end{document}